\documentclass[fleqn,usenatbib]{mnras}

\usepackage{newtxtext,newtxmath}

\usepackage[T1]{fontenc}

\DeclareRobustCommand{\VAN}[3]{#2}
\let\VANthebibliography\thebibliography
\def\thebibliography{\DeclareRobustCommand{\VAN}[3]{##3}\VANthebibliography}

\usepackage{graphicx}	
\usepackage{amsmath}	

\title[Discovery of two new GW Vir stars]{Pulsating hydrogen-deficient white dwarfs and pre-white dwarfs observed with  TESS -- IV. Discovery of two new GW Vir stars: TIC\,0403800675 and TIC\,1989122424}

\author[Murat Uzundag et al.]{
Murat Uzundag,$^{1,2}$\thanks{E-mail: murat.uzundag@postgrado.uv.cl}, Alejandro H. C\'orsico$^{3,4}$, S. O. Kepler$^{5}$, Leandro G. Althaus$^{3,4}$, Klaus Werner$^{6}$, \newauthor Nicole Reindl$^{7}$ and Maja Vu\v{c}kovi\'{c}$^{1}$
\\
$^{1}$Instituto de F\'isica y Astronom\'ia, Universidad de Valpara\'iso, Gran Breta\~na 1111, Playa Ancha, Valpara\'iso 2360102, Chile
           \\
           $^{2}$European Southern Observatory, Alonso de Cordova 3107, Santiago, Chile
           \\
           $^{3}$Grupo de Evoluci\'on Estelar y Pulsaciones. 
           Facultad de Ciencias Astron\'omicas y Geof\'{\i}sicas,
           Universidad Nacional de La Plata, Paseo del Bosque s/n, 1900, Argentina
           \\
           $^{4}$IALP - CONICET
           \\
           $^{5}$Instituto de F\'{i}sica, Universidade Federal do Rio Grande do Sul, 91501-970, Porto-Alegre, RS, Brazil
           \\
           $^{6}$Institut f\"ur Astronomie und Astrophysik, Kepler Center for Astro and Particle Physics, Eberhard Karls Universit\"at, Sand 1,72076 T\"ubingen, Germany
           \\
           $^{7}$Institute for Physics and Astronomy, University of Potsdam, Karl-Liebknecht-Str. 24/25, D-14476 Potsdam, Germany
}

\date{Accepted XXX. Received YYY; in original form ZZZ}

\pubyear{2022}

\begin{document}
\label{firstpage}
\pagerange{\pageref{firstpage}--\pageref{lastpage}}
\maketitle

\begin{abstract}

We present two new GW Vir-type pulsating white dwarf stars, TIC\,0403800675 (WD\,J115727.68-280349.64) and TIC\,1989122424 (WD J211738.38-552801.18) discovered in the Transiting Exoplanet Survey Satellite (TESS) photometric data. 
For both stars, the TESS light curves reveal the presence of oscillations with periods in a narrow range 
between 400 and 410\,s, which are associated with typical gravity ($g$)-modes.
Follow-up ground-based spectroscopy shows that both stars have similar effective temperature ($T_\mathrm{eff} = 110,000 \pm 10,000$\,K) and surface gravity ($\log g = 7.5 \pm 0.5$), but different He/C composition (mass fractions): He\,=\,0.75 and C\,=\,0.25 for TIC\,0403800675, and He\,=\,0.50  and C\,=\,0.50 for TIC\,1989122424. By performing a fit to their spectral energy distributions, we found for both stars radii and luminosities of $R=0.019\pm0.002\,R_\odot$ and $\log(L/L_\odot)=1.68^{+0.15}_{-0.24}$, respectively. 
By employing evolutionary tracks of PG~1159 stars, we find the masses of both stars to be $0.56\pm0.18  M_{\odot}$ from the  $\log g$-$T_\mathrm{eff}$ diagram and  $0.60^{+0.11}_{-0.09} M_{\odot}$ from the Hertzsprung Russell diagram.


\end{abstract}




\section{Introduction}
\label{intro}

White dwarf (WD) stars are the end evolutionary state of all stars formed with 
initial masses below around 7$-11 M_{\odot}$, which comprise more than 95\% of all 
stars in our galaxy \citep{2010A&ARv..18..471A}. In the course of their evolution,  
WDs cross at least one phase of pulsational instability that converts them into 
pulsating variable stars. GW Vir variable stars are the hottest known 
class of pulsating WDs and pre-WDs, with 75\,000 K $\leq T_{\rm eff} \leq$ 250\,000 K 
and 5.3 $\leq$ log$g$ $\leq$ 8 \citep{2021arXiv211113549W}. They are located within a definite
instability strip \citep[see Fig. 1 of][]{2019A&ARv..27....7C}.  GW~Vir stars include 
some objects that are still surrounded by a nebula, called the variable planetary nebula nuclei (PNNVs), and some objects that lack a nebula, which are called DOVs. Both groups (DOVs and 
PNNVs) are frequently referred to as GW~Vir variable stars. GW~Vir stars display brightness fluctuations with periods in the range $300 - 6\,000$ s, and amplitudes up to a few mmag (1~mmag=1~ppt), associated  with low-order ($\ell \leq 2$) non-radial $g$(gravity) modes. 

GW~Vir stars are pulsating PG~1159 stars (after the prototype of the spectroscopic class, the star PG~1159$-$035), which are hydrogen(H)-deficient post-AGB stars with surface layers rich in helium (He), carbon (C) and oxygen (O) \citep{2006PASP..118..183W}.
It is believed that this mixture is a result of a mixing event produced by a late He flash during the so-called born-again episode  \citep{1977PASJ...29..331F,1979A&A....79..108S, 1983ApJ...264..605I, 2010A&ARv..18..471A}. PG~1159 stars are considered the evolutionary link between post-AGB stars and most of the H-deficient WDs, including DO 
and DB WDs  \citep{1999A&A...349L...5H,2005A&A...435..631A,  Sowicka2021}. 
They can be either the outcome of single star evolution (late thermal pulse scenario, LTP, or very late thermal pulse scenario, VLTP) or binary star evolution (double WD merger). The classification of GW Vir stars includes also the pulsating Wolf-Rayet central stars of a planetary nebula ([WC]) and Early-[WC] = [WCE] stars since they share the same pulsation properties of pulsating PG~1159 stars \citep{2007ApJS..171..219Q}.  
Thus far about 50 PG~1159 stars have been identified \citep{2021arXiv211113549W}.
Amongst them, approximately 50\% \citep[22 objects; see][]{2019A&ARv..27....7C,2021A&A...655A..27U}  
have been discovered to be pulsating. It is especially important to 
find new pulsators of this class, as they can provide insight into the AGB and VLTP/LTP 
phases, as well as angular momentum loss throughout the extensive mass loss phases \citep[and references therein]{Kepler2014}.


Prior to space missions, GW~Vir stars were monitored through long-term observations carried out by the multisite photometric campaign with the ''Whole Earth  Telescope'' \citep[WET;][]{1990ApJ...361..309N}. 
These observations provided invaluable sources of information to constrain their internal structure \citep{1991ApJ...378..326W}. 
The spectral observations from the Sloan Digital Sky Survey
\citep[SDSS,][]{2000AJ....120.1579Y} promoted the discovery of a GW~Vir pulsating star, SDSS~J075415.12+085232.18 \citep{Kepler2014}. The advent of the {\it Kepler} space mission has resulted in important advances in the study of pulsating stars in general \citep{2021RvMP...93a5001A,2022arXiv220111629K} and pulsating WDs in particular \citep{2020FrASS...7...47C}. Unfortunately, during the main mission of 
the {\it Kepler} satellite \citep{2010Sci...327..977B}, no GW~Vir star was observed.
 During the {\it Kepler} extended mission \citep[{\it K2};][]{2014PASP..126..398H}, the prototype PG\,1159$-$035 was observed during almost 50 days of coverage and the findings will be reported soon (G. O. da Rosa et al. in preparation). 
Currently, uninterrupted observations from space with the Transiting Exoplanet Survey Satellite (TESS) allow us to find and characterize new and already known GW Vir stars. Indeed, TESS has allowed a detailed asteroseismological analysis of a number of formerly known GW~Vir stars \citep{2021A&A...645A.117C}, 
enabling the determination of their fundamental parameters and evolutionary properties. The discovery of two new GW~Vir stars has been presented by \citet{2021A&A...655A..27U}. In these studies, using asteroseismic techniques (e.g., asymptotic period spacing and rotational splittings), the authors were able to determine the internal chemical stratification, total mass and, in some cases, rotation velocity of GW Vir stars.

In this work, we present the discovery of two new GW~Vir stars, TIC\,0403800675 (WD\,J115727.68-280349.6) and TIC\,198912242 (WD\,J211738.38-552801.1), which were observed during the survey phase of 
the southern ecliptic hemisphere cycle 1 and 3 of TESS. 
In addition, for each target we obtained low-resolution spectra and fitted model 
atmospheres to estimate their fundamental atmospheric parameters, and 
examined the TESS light curve to identify the pulsational modes. 
This study is the fourth part of a series of papers devoted to the study of 
pulsating H-deficient WDs and pre-WDs observed with TESS. 
The first article was devoted to a set of six already known GW~Vir stars including PNNVs and DOVs  \citep{2021A&A...645A.117C},
the second to the discovery of two new GW~Vir stars of the DOV type \citep{2021A&A...655A..27U}, and the third to  a detailed asteroseismological analysis of the prototype of the pulsating DB WD, GD~358 \citep{2022A&A...659A..30C}. 

The paper is organized as follows.
In Sect. \ref{spectroscopy}, we present the details of spectroscopic observations and the data reduction. 
In  Sect. \ref{spec_fit}, we derive atmospheric parameters for each star by fitting synthetic spectra to the newly obtained low-resolution spectra.
In  Sect. \ref{photometry}, we analyze the photometric TESS data and give details on the frequency analysis. Finally, in Sect. \ref{conclusion}, we  summarize  our  main  results.

\section{Spectroscopy}
\label{spectroscopy}

TIC\,0403800675 (WD\,J115727.68$-$280349.6) and TIC\,198912242 (WD\,J211738.38$-$552801.1) were classified as WD candidates by \citet{GentileFusillo19} from their colors and Gaia DR2 parallax.
The Gaia DR3 parallax and corresponding distance for TIC\,0403800675 are  $\pi= 1.86^{+0.07}_{-0.06}$\,mas and $d = 535.41^{+19.49}_{-18.44}$\,pc, while for TIC\,198912242 are  $\pi= 1.45^{+0.05}_{-0.06}$\,mas and $d = 688.27^{+22.34}_{-26.31}$\,pc \citep{Bailer-Jones2021}, respectively.



To estimate the atmospheric parameters of TIC\,0403800675 and TIC\,1989122424, we obtained 
spectra with the Goodman High-Throughput Spectrograph \citep[GHTS,][]{clemens2004} at the SOAR 4.1-m telescope on Cerro Pach\'on. 
We reduced the spectroscopic data using the instrument pipeline\footnote{\url{https://github.com/soar-telescope/goodman_pipeline}} including overscan, trim, slit trim, bias and flat corrections. 
We employed a method developed by \citet{wojtek2004}, which is included in the pipeline, to identify and remove cosmic rays.
After the reduction was completed, the wavelength  calibration has been applied by using \textit{PyRAF}\footnote{\url{http://www.stsci.edu/institute/software_hardware/pyraf}} \citep{pyraf2012}. 
We used the frames produced with the internal He-Ar-Ne comparison lamp at the same telescope position as the targets in order to apply wavelength calibrations.
The sixth order Legendre function is used to calibrate the pixel-wavelength correspondence using an atlas of known He-Ar-Ne lines.
 Finally, we used the standard star Feige\,110 seen with the identical apparatus to normalize the spectra with a high-order Legendre function.
Table \ref{table:spectroscopy} contains the details of the spectroscopic observations, including the name of the targets, right ascension, declination, Gaia magnitude, date, exposure time, grating, resolution, and S/N ratio.

\begin{table*}
\renewcommand{\arraystretch}{1.2}
\caption{Log of spectroscopic observations.}
\begin{tabular}{cccccccccc}
\hline \hline
TIC & Name & RA & Dec  & $G_{mag}$ & Obs. Date & Exp.  & Grating & Resolution  & S/N \\
&  & (J2000) &  (J2000)& & (UT)       & (sec)  & (l mm$^{-1}$) & ($\Delta \lambda$ (\AA)) &  \\
\hline                                           
0403800675 & WD~J115727.68$-$280349.64  & 11:57:28 & -28:03:53 & 16.16 & 2021-06-18 03:26:17 & 1200 & 400 & 4.6 &  65  \\
1989122424 & WD~J211738.38$-$552801.18  & 21:17:40 & -55:28:16 & 16.75 & 2021-06-19 07:54:54 &  900 & 400 & 4.6 &  70  \\

\hline 
\label{table:spectroscopy}
\end{tabular}
\end{table*}

\section{Spectral fitting}
\label{spec_fit}

The spectral lines in the spectra of TIC\,0403800675 and TIC\,1989122424 are all from He\,II and C\,IV. 
Oxygen, which is the most abundant element in PG\,1159 stars after He and C, and nitrogen (N) ---which is present as a trace element in some PG\,1159 stars--- may be visible in spectra with higher resolution and signal-to-noise ratio.
In the spectra of both stars, there are no indications of the presence of H.

For the spectral analysis, we used a grid of line-blanketed non-local thermodynamic equilibrium (non-LTE) model atmospheres consisting of H, He, and C as introduced by \citet{2014A&A...564A..53W}. 
The grid spans $T_\mathrm{eff}$ = 60,000--140,000\,K in effective temperature and $\log g$ = 4.8--8.3 in surface gravity, with steps of 5,000\,K or 10,000\,K and 0.3\,dex, respectively. C/He mass ratios in the range 0.0--1.0 were considered, namely C/He = 0.0, 0.03, 0.09, 0.33, 0.77, and 1.0.   The hydrogen abundance was set to zero. Synthetic spectra were convolved with a Gaussian accounting for the spectral resolution of the observations. The best fitting models were chosen by visual comparison with the rectified observed spectra. 

\begin{figure*}
	\includegraphics[width=0.9\textwidth]{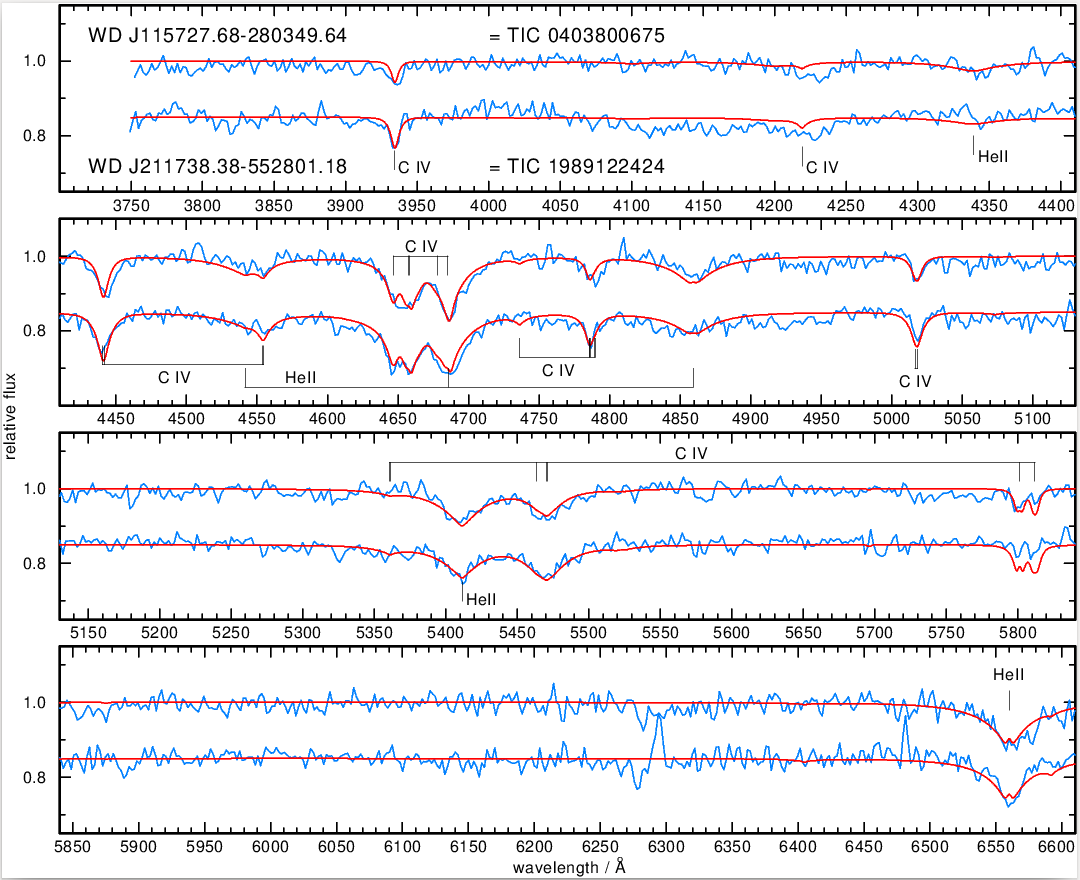}
    \caption{Optical spectra of the two new GW Vir stars (blue graphs) obtained with SOAR/Goodman. Overplotted are the best-fit models (red). Identifications of He\,II and C\,IV lines are marked.} 
\label{fig:spectral_fits}
\end{figure*}

The model fits are depicted in Fig.\,\ref{fig:spectral_fits}. Both stars have $T_\mathrm{eff} = 110,000 \pm 10,000$\,K and $\log g = 7.5\pm0.5$, but a different atmospheric composition. For  TIC\,0403800675, we found ${\rm He} = 0.75^{+0.05}_{-0.15}$  and ${\rm C}= 0.25^{+0.15}_{-0.05}$ and for TIC\,1989122424, we measured ${\rm He}= 0.50^{+0.20}_{-0.05}$ and ${\rm C}= 0.50^{+0.05}_{-0.20}$ (mass fractions). 
 An upper limit to the abundance of hydrogen was determined to be 5\% by mass.
At this abundance, a H$\alpha$ emission core would be detectable and the H$\beta$ line blend with the respective He\,II line would be too strong.

\begin{figure*}
	\includegraphics[width=0.75\textwidth]{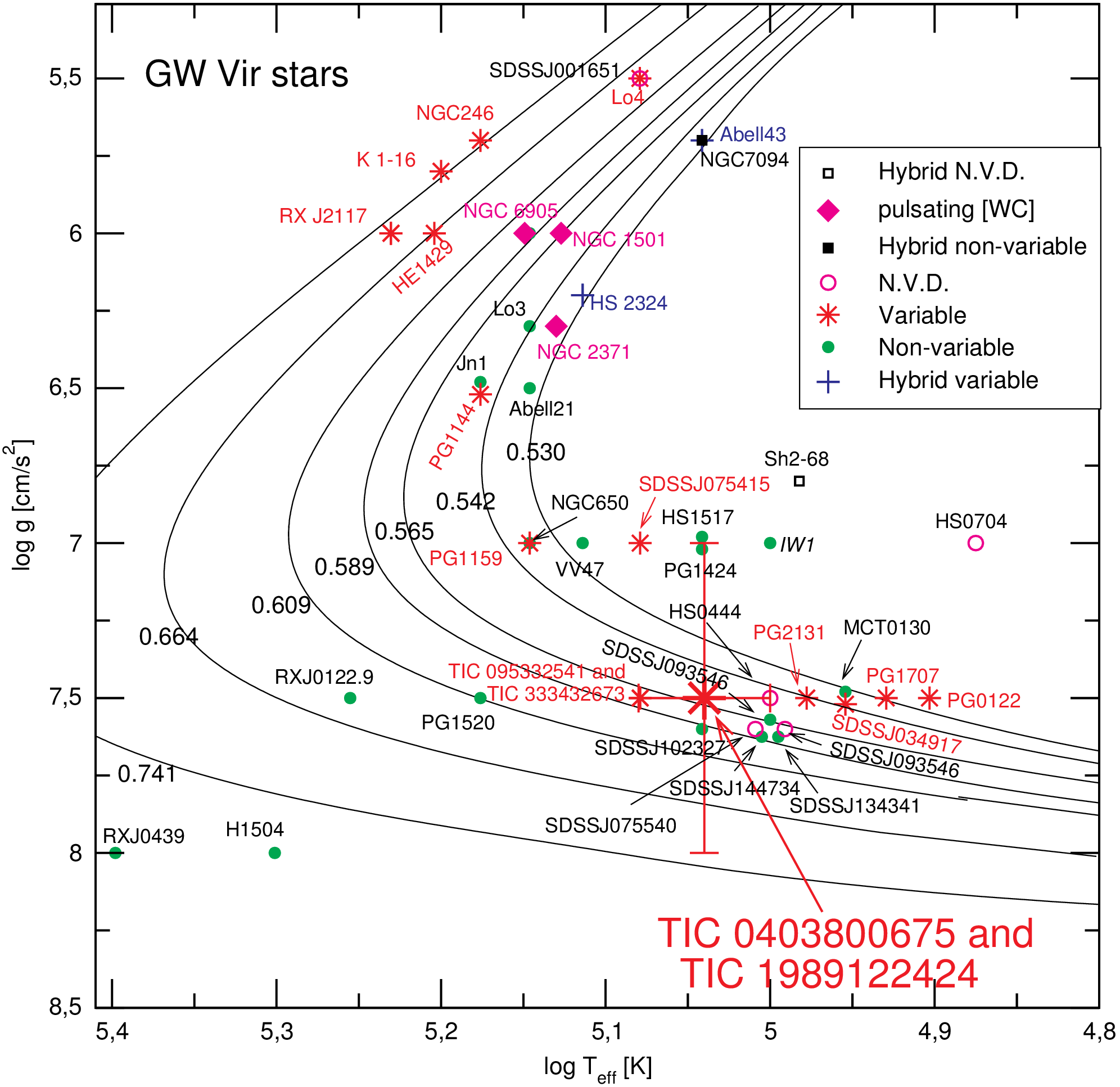}
    \caption{
    The already known variable and non-variable PG\,1159 stars and variable [WCE] stars in the 
    $\log T_{\rm eff} - \log g$ diagram.
    Thin solid black curves show the post$-$born again evolutionary tracks from \citet{2005A&A...435..631A} and  \citet{2006A&A...454..845M} for different stellar masses noted on top of each track.
    "N.V.D." stands for PG~1159 stars  with no variability data. "Hybrid" refers to PG~1159 stars exhibiting H in their atmospheres. The location of the two new GW Vir stars  TIC\,0403800675 and TIC\,1989122424 is emphasized with a large red star symbol and error bars. 
    Both stars share the same spectroscopic surface parameters, 
    $T_{\rm eff} = 110\,000 \pm 10\,000$ K and $\log g = 7.5 \pm$ 0.5. }
    \label{fig:logTefflogg}
\end{figure*}

In Fig. \ref{fig:logTefflogg} we show the location of the new GW~Vir stars in the  $\log g$-$T_\mathrm{eff}$ diagram. 
By linear interpolation among the PG~1159 evolutionary tracks of \cite{2005A&A...435..631A} and \cite{2006A&A...454..845M}, we derive a stellar mass of $M_{\star}= 0.56\pm0.18 M_{\sun}$ for TIC\,0403800675 and TIC\,1989122424.  These evolutionary tracks of PG~1159 stars have been compared by \cite{2007A&A...470..675M} with other independent published calculations.   
These authors concluded that the differences in the inferred masses using 
different sets of PG1159 evolutionary tracks must be less than
$\sim 0.01 M_{\odot}$. This constitutes the systematic errors associated with the
\cite{2005A&A...435..631A} and \cite{2006A&A...454..845M} evolutionary tracks. We conclude that the errors in the spectroscopic mass of the two new GW Vir stars 
come mainly from the uncertainties in the spectroscopic $\log g$ and $T_{\rm eff}$.

\section{SED fitting}
\label{sed_fit}

In order to determine the radii of the two stars we performed a fit to the
spectral energy distribution (SED), by varying the solid angle $\pi(R/d)^2$,
(which relates the flux at the surface of the system to what is received at
Earth) until a good agreement of the  predicted fluxes and the
observations was found as shown in Fig.\ref{sed_fit}. We employed our best fitting model atmospheres for both stars, and using the \cite{Fitzpatrick1999} reddening 
law, our  predicted spectra were reddened for different values of $E_\mathrm{B-V}$.
We used the distance provided  by \cite{Bailer-Jones2021} and 
employed photometry from  GALEX \citep{Bianchi2014}, 
Gaia DR2 and eDR3 \citep{2021A&A...650C...3G}, and 
Pan-STARRS1 \citep{Chambers+2016}. Magnitudes were converted into fluxes using 
the VizieR Photometry viewer\footnote{\url{http://vizier.unistra.fr/vizier/sed/}}.
For TIC\,0403800675 we find a reddening of $E_\mathrm{B-V}=0.034$\,mag and for 
TIC\,1989122424 we find $E_\mathrm{B-V}=0.048$\,mag. Both values are in agreement 
with the upper limits of the 2D dust map provided by \cite{SchlaflyFinkbeiner2011}. 
For both stars we determine a radius of $R=0.019\pm0.002\,R_\odot$. Using 
$L=4\pi\,\sigma R^2 T_\mathrm{eff}^4$, where $\sigma$ is the Stefan-Boltzmann 
constant, we calculate for both stars a luminosity of 
$\log(L/L_\odot)=1.68^{+0.15}_{-0.24}$.
This allows us now to also derive the masses for TIC\,0403800675 and TIC\,1989122424 
in the  Hertzsprung–Russell diagram (HRD,  see Fig. \ref{hrd_fig}). By linear interpolation among the PG~1159 evolutionary tracks of \cite{2006A&A...454..845M}, we derive a stellar mass of 
$M_{\star}= 0.60^{+0.11}_{-0.09} M_{\sun}$ for both stars, which is in agreement 
with the masses derived from the $\log g$-$T_\mathrm{eff}$ diagram (see Fig.\ref{spec_fit}).
The main characteristics of TIC\,0403800675 and TIC\,1989122424 are listed in Table~\ref{table:all_param} including atmospheric parameters, masses, luminosities, radii, distances and reddening. 

\begin{figure}
	\includegraphics[width=\columnwidth]{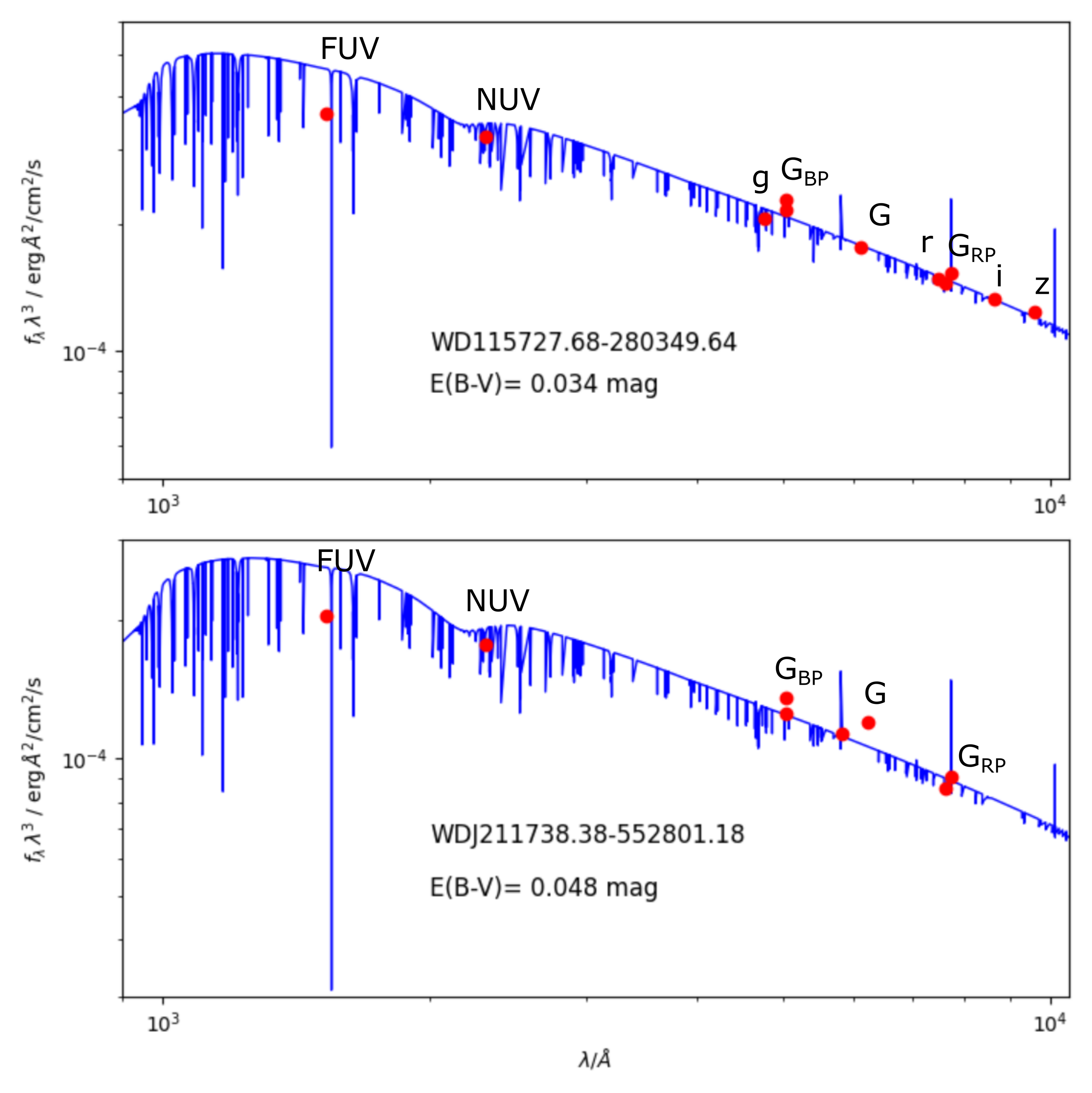}
    \caption{Fit to the SEDs of TIC\,0403800675 (WD\,J115727.68-280349.64, upper panel) and TIC\,1989122424 (WD\,J211738.38-552801.18, bottom panel). The red dots indicate filter-averaged fluxes that were converted from observed magnitudes. Error bars are smaller than symbol size. The blue solid lines represent our best fitting model fluxes with the parameters as stated in Section \ref{sed_fit}.
    The assumed values for the interstellar reddening in each fit are indicated.}
    \label{sed_fig}
\end{figure}

\begin{figure}
	\includegraphics[width=\columnwidth]{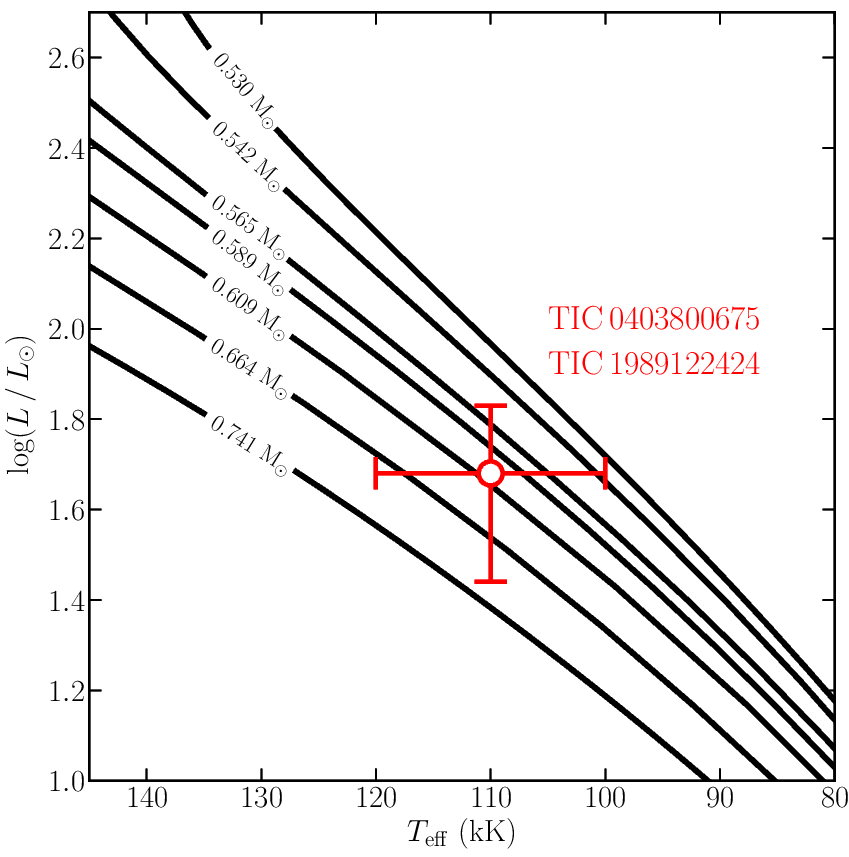}
    \caption{The location of the two new GW Vir stars TIC\,0403800675 and TIC\,1989122424 (emphasized with a large red dot symbol and error bars) in the HRD. Black curves show the post$-$born again evolutionary tracks from \citet{2006A&A...454..845M} for different stellar masses. Both stars share the same effective temperatures ($T_{\rm eff} = 110\,000 \pm 10\,000$ K) and luminosities ($\log(L_{\star}/L_{\odot})=1.68^{+0.15}_{-0.24}$).}
    \label{hrd_fig}
\end{figure}

\begin{table}
\centering
\caption{Properties of GW Vir pulsating stars studied in this work.} 
\begin{tabular}{l|cc}
\hline
\hline

Quantity & TIC\,0403800675 &  TIC\,1989122424 \\ 
\hline
$T_{\rm eff}$ [kK]                            &    $110 \pm 10$  &   $110 \pm 10$      \\
$M_{\star}$ [$M_{\odot}$]                     &  $0.56\pm0.18$  &   $0.56\pm0.18$       \\ 
$\log g$ [cm/s$^2$]                           &    $7.5\pm0.5$     &   $7.5\pm0.5$       \\ 
$({\rm He},{\rm C}) $                         & $0.75^{+0.05}_{-0.15}$, $0.25^{+0.15}_{-0.05}$  &   $0.50^{+0.20}_{-0.05}$, $0.50^{+0.05}_{-0.20}$  \\ 
$R_{\star}$ [$R_{\odot}$]                     &    $0.019\pm0.002$  &     $0.019\pm0.002$     \\
$\log (L_{\star}/L_{\odot})$                  &   $1.68^{+0.15}_{-0.24}$       &    $1.68^{+0.15}_{-0.24}$  \\ 
$E(B-V)$ [mag]                                     &    0.034      &     0.048     \\
$\pi$ [mas]                                   &    $1.86^{+0.07}_{-0.06}$    &   $1.45^{+0.05}_{-0.06}$  \\ 
$d$  [pc]                                     &   $535.41^{+19.49}_{-18.44}$  &   $688.27^{+22.34}_{-26.31}$  \\ 

\hline
\hline
\end{tabular}
\label{table:all_param}
\end{table}

\section{Photometric observations --- TESS}
\label{photometry}

TIC\,0403800675 was observed  by TESS in sector 10 between 26 March to 22 April 2019 and in sector 36 between 7 March and 2 April 2021. 
TIC\,1989122424 was observed in a single sector 27 between 4 July and 30 July 2020 with only 120\,sec cadence. 

The light curves were downloaded from The Mikulski Archive for Space Telescopes, which is hosted by the Space Telescope Science Institute
(STScI)\footnote{\url{http://archive.stsci.edu/}} as FITS format.  
The light curves were processed by the Science Processing Operations Center (SPOC) pipeline \citep{Jenkins2016}.
We first downloaded the target pixel file (TPF) of interest from the MAST archive, which is maintained by the Lightkurve Collaboration; \citet{lightkurve2020}. 
The TPFs comprises an 11x11 postage stamp of pixels from the one of four CCDs per camera that the target is located on. 
The TPFs are examined to determine the amount of crowding and other potential bright sources near the target.  
Because both targets had a modest amount of crowding which we evaluated using the $\tt CROWDSAP$ parameter, that was set to 0.6 for TIC\,1989122424 and 0.8 for TIC\,0403800675.
Therefore we have decided to use the pipeline aperture as it gave the most optimal result with respect to signal-to-noise ratio.
We extracted times in barycentric corrected dynamical Julian days and fluxes (PDCSAP FLUX) from the FITS files. We converted the fluxes to fractional variations from the mean and transformed to amplitudes in parts-per-thousand (ppt). 
Finally, the data were sigma-clipped based on 5~$\sigma$ to remove the outliers which appear above 5 times the median of intensities.
The resulting short-cadence (SC) light curve of TIC\,040380067 comprises 29\,298 images spanning 47.4 days, 
while the resulting SC light curve of TIC\,1989122424 includes 16\,478 data points spanning 23.9 days. 
TIC\,0403800675 was also observed with ultra-short-cadence (USC) during 23.9 days in sector 36 that yield 93\,122 images.


\subsection{Frequency solution}

The Fourier transform (FT) was  used to examine the periodicities present in the light curves. 
The frequencies, amplitudes and errors were estimated using both the ${\tt Period04}$ and our custom tool.  
We fitted each frequency that appears above the 0.1\%\ false alarm probability (FAP).
The FAP level was calculated by reshuffling the light curves 1000 times as described in \citep{Kepler1993}. 
The errors of each frequency and amplitude of the pulsation modes were
estimated using Monte Carlo simulations \citep{LB2005}. 
Furthermore, we have calculated sliding Fourier transform (sFT) for both stars in order to see the temporal evolution of the pulsational modes over the course of TESS observations. 
To do so, we use a nine-day sliding window with a 2-day step size. 
A color-scale in ppt units is used to depict the amplitudes.
Following that, we calculate the Fourier transform of each subset and trail them in time.

\begin{figure}
	\includegraphics[width=0.47\textwidth]{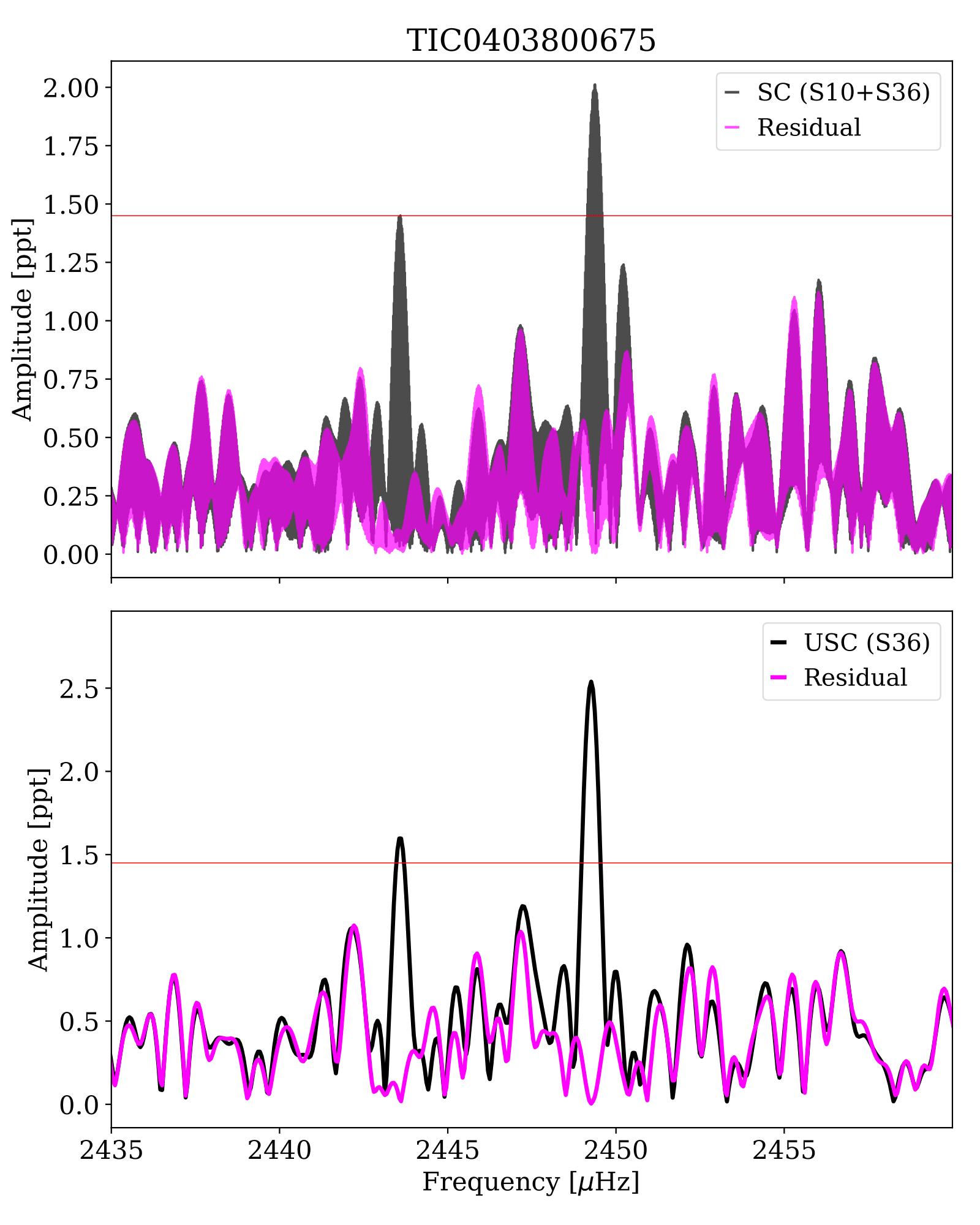}
	\includegraphics[width=0.49\textwidth]{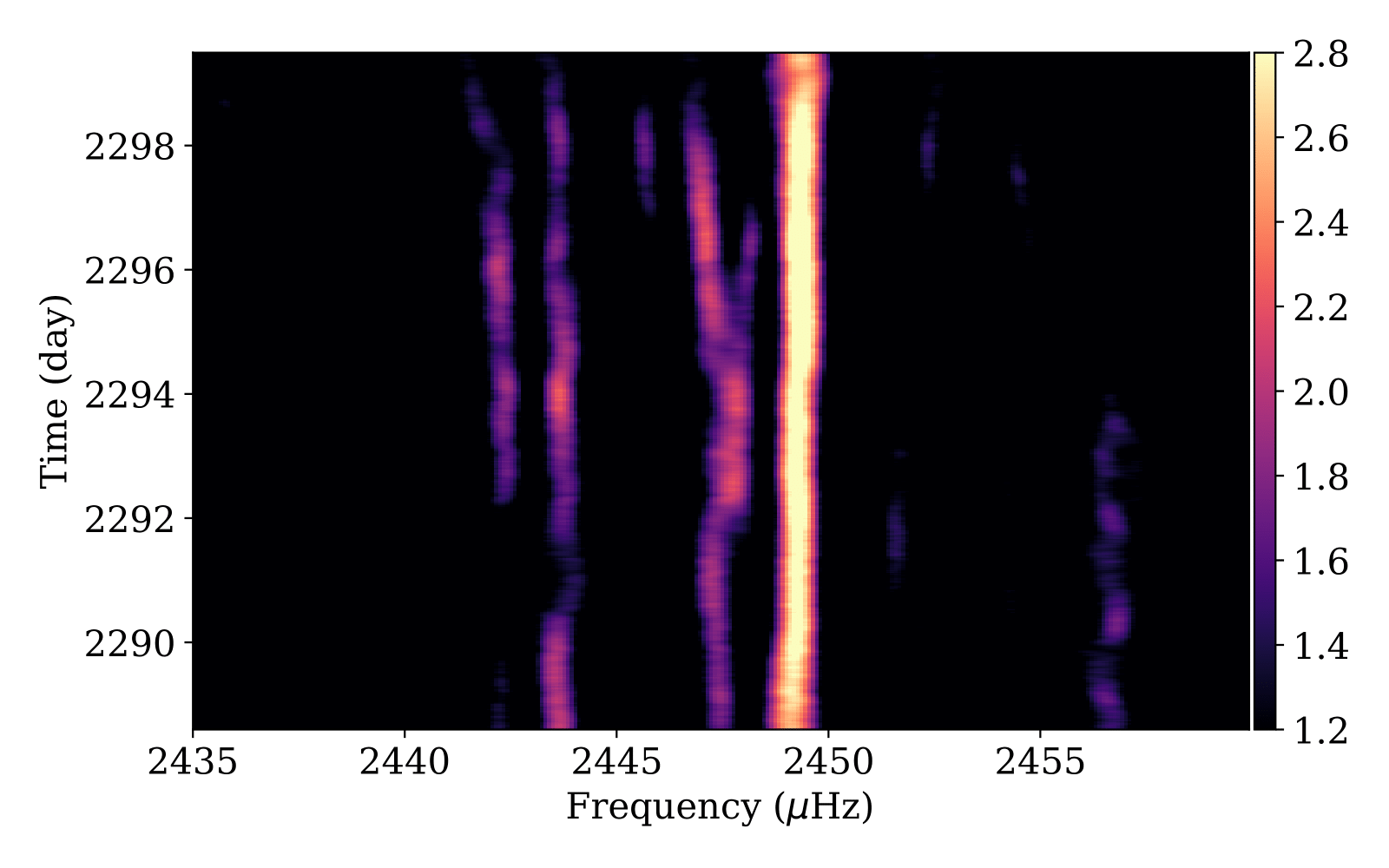}
    \caption{{\sc Top:} Fourier transform of short-cadence data of sector 10 and 36 (black lines)  for TIC\,0403800675. The magenta line depicts the FT of the prewhitened light curve. The horizontal red line indicates the 0.1\% FAP level. 
   {\sc Middle:} Fourier transform of ultra-short-cadence data of sector 36 (black lines) for TIC\,0403800675. The red and magenta lines are the same as top panel the 0.1\% FAP level and the FT of the prewhitened light curve, respectively. 
  {\sc Bottom:}  Sliding Fourier transform of ultra-short-cadence data of sector 36 data of TIC\,0403800675. The color-scale illustrates amplitude in  parts-per-thousand.}
    \label{FT04}
\end{figure}

For TIC\,0403800675, we analyzed TESS observations using the SC and USC modes. The SC mode samples every 2-minutes allowing us to analyze the frequency range up to the Nyquist frequency at about 4167 $\mu$Hz, while the USC mode samples every 20-seconds permitting us to examine the frequency range up to the Nyquist frequency at about 25\,000 $\mu$Hz.
For TIC\,0403800675, we detected two significant frequencies at 2445 $\mu$Hz and 2450 $\mu$Hz. In Fig.\ref{FT04}, we show FT of combined SC data of sector 10 and 36 (black lines) (top panel), while the middle panel, we present USC data of sector 36. The magenta lines show the FT of the prewhitened light curves.  The signals are 20\% more amplified in USC data. Sliding Fourier transform display the both identified peaks in Fig.\ref{FT04}. 
The frequencies detected for TIC\,0403800675 are presented in Table \ref{tab:TIC04}. 

\begin{figure}
	\includegraphics[width=0.47\textwidth]{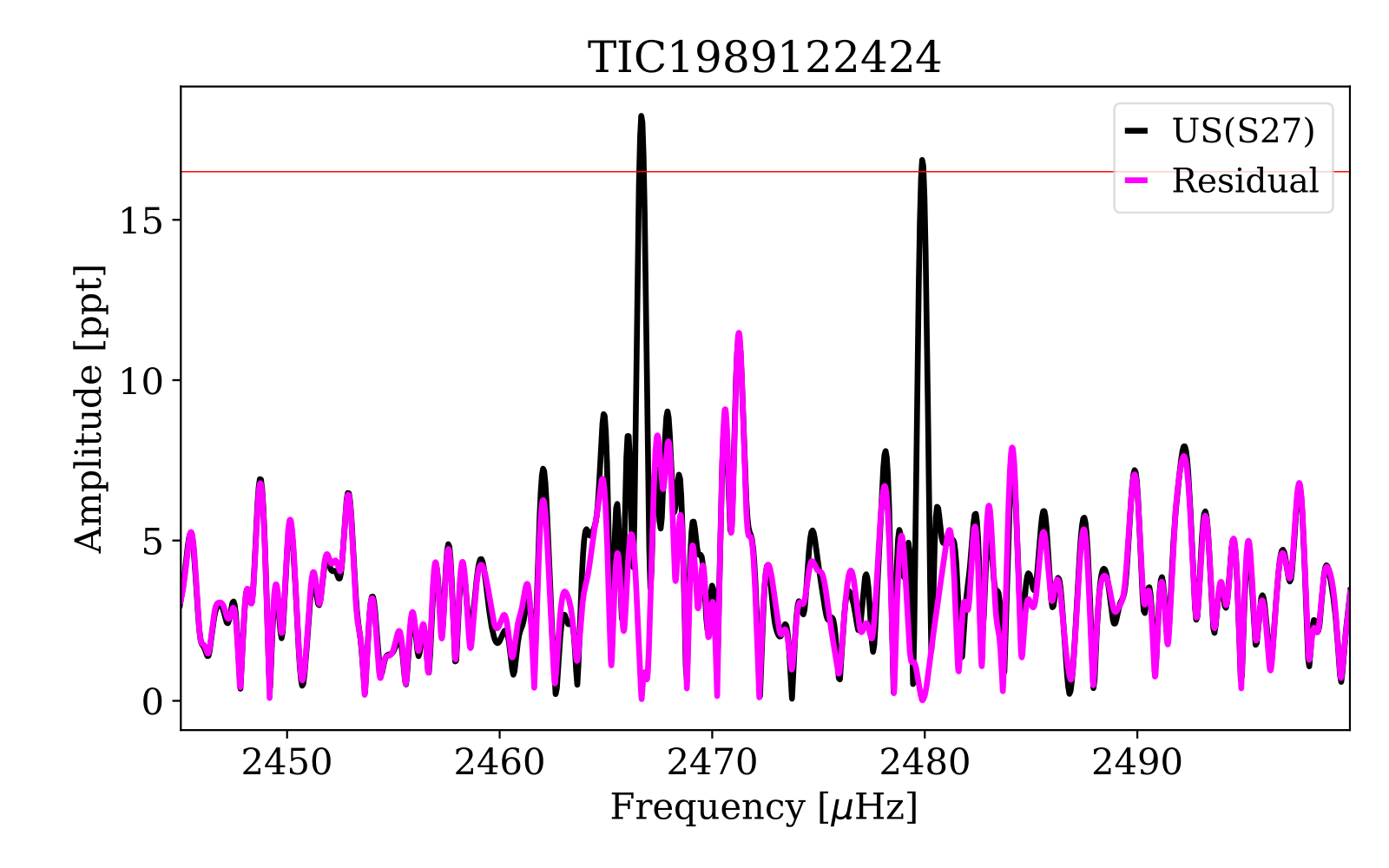}
	\includegraphics[width=0.49\textwidth]{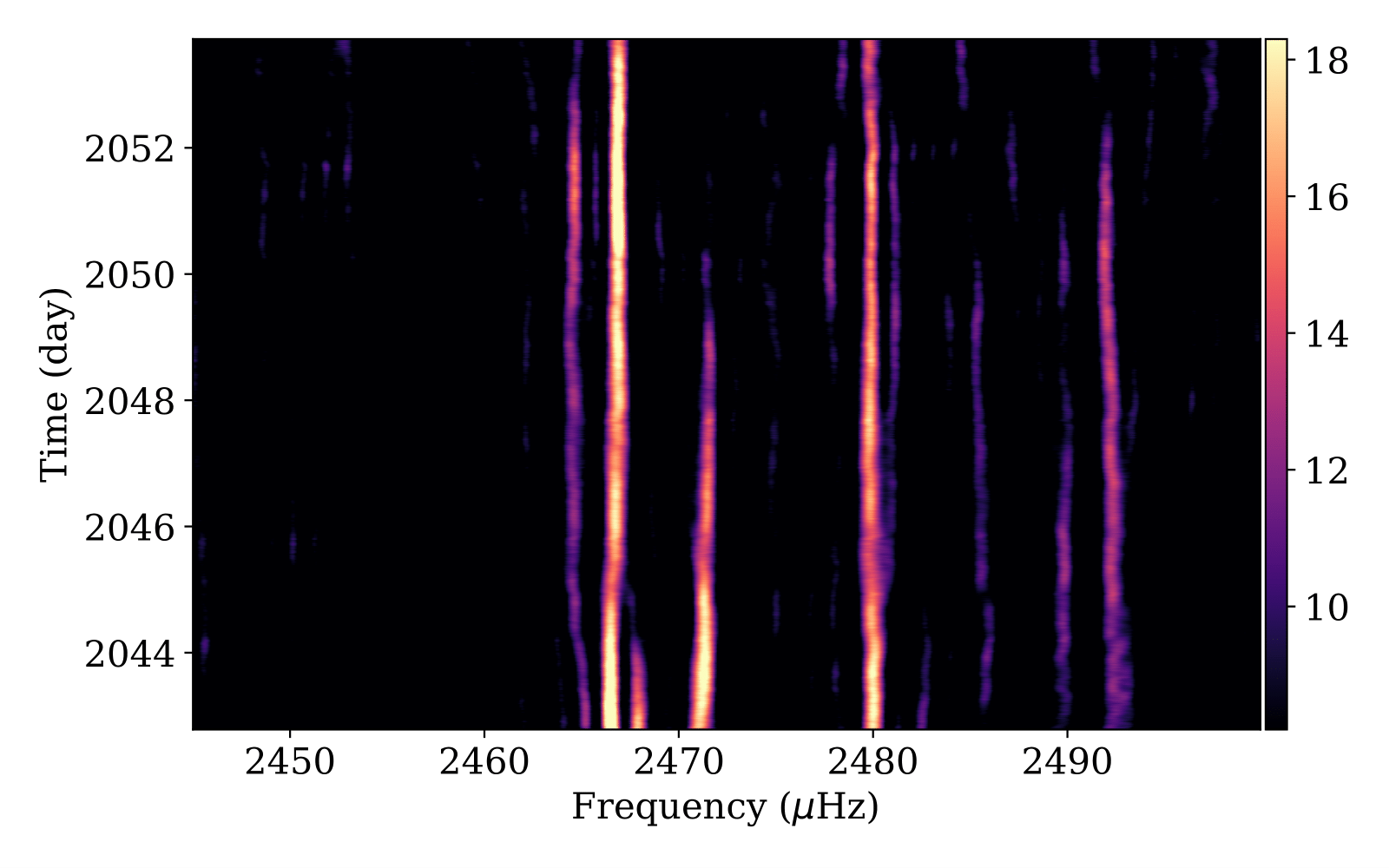}
    \caption{{\sc Top:} Fourier transform of short-cadence data of sector 27 of TIC\,1989122424. The horizontal red line indicates the 0.1\% FAP level. The magenta line depicts the FT of the prewhitened light curve.
 {\sc Bottom:}  Sliding Fourier transform of the same data-set of TIC\,1989122424. The color-scale illustrates amplitude in  parts-per-thousand.}
    \label{FT198}
\end{figure}

For TIC\,1989122424, we also detected two clear signals at similar region between 2450 $\mu$Hz and 2490 $\mu$Hz using solely SC observations. In Fig.\ref{FT198}, we focus on this region and show  the FT of the original light curve (black line) and the FT of the prewhitened light curves (magenta line). In sFT, these peaks also clearly visible along with some other peaks. 
The two reported peaks at 2466 $\mu$Hz and 2479 $\mu$Hz are clearly visible in sFT and they are stable during the run. 
We also see that there is a peak at 2470 $\mu$Hz, which is stable during the first ten days. For this peak, we did not produce an  non-linear least square (NLLS) fit to extract from the light curve. We need additional observations to confirm whether it is real pulsational peak or not. 
The detected frequencies for TIC\,0403800675 are reported in Table \ref{tab:TIC198}. 

We hypothesized that the two closely spaced frequencies in both amplitude spectra and sFT can be due to stellar rotation in which the pulsation frequencies are split into 2$\ell +$1 azimuthal components. 
Assuming these frequencies are dipole modes and result from stellar rotation, then the rotation periods would range from 1.02 to 2.04 d (depending on the missing azimuthal order) for TIC\,0403800675 and 0.43 d to 0.87 d for TIC\,1989122424. 
Given the rotation periods of the GW Vir pulsating stars, which range from 5 hours to a few days \citep{2019A&ARv..27....7C,2021A&A...645A.117C}, any of the potential solutions that we estimate for TIC\,0403800675 and TIC\,1989122424 could be conceivable.
Furthermore, we note that the rotational periods of GW Vir stars cover about the same period range as white dwarfs showing ultra-highly excited (UHE) metals in their optical spectra. \cite{Reindl+2021} discovered recently that this latter class of white dwarfs show photometric periods between 6\,hours to 3\, days and that their variability is likely caused by spots on the surfaces of these stars and/or geometrical effects of circumstellar material (see also \citealt{Reindl+2019}).
Since UHE white dwarfs are considered to be in an evolutionary stage immediately following that of the GW Vir pulsators, it would be interesting to know the rotational periods of more GW Vir stars in order to test a possible evolutionary connection. 

\begin{table}
	\centering
	\caption{Identified frequencies, periods, and amplitudes (and their uncertainties) and the signal-to-noise ratio in the  data of TIC\,0403800675. 
Frequency and amplitude that are detected in both sector 10 and 36.
}
	\label{tab:TIC04}
	\begin{tabular}{rcccc} 
\hline
Peak & $\nu$    &  $\Pi$  &  $A$   &  S/N  \\
 & ($\mu$Hz)      &  (s)   & (ppt)   &   \\
 \hline 
f$_{\rm 1}$ &  2443.597(51)  & 409.232(86) &  1.63(33) &	4.3   \\
f$_{\rm 2}$ &  2449.261(33)  & 408.286(55) &  2.54(33) &	6.3   \\

\hline 
\end{tabular}
\end{table}

\begin{table}
	\centering
	\caption{Identified frequencies, periods, and amplitudes (and their uncertainties) and the signal-to-noise ratio in the  data of TIC\,1989122424. 
}
	\label{tab:TIC198}
	\begin{tabular}{rcccc} 
\hline
Peak & $\nu$    &  $\Pi$  &  $A$   &  S/N  \\
 & ($\mu$Hz)      &  (s)   & (ppt)   &   \\
 \hline 
f$_{\rm 1}$ &  2466.673(41)  & 405.404(68) &  18.35(2.89) &	5.1   \\
f$_{\rm 2}$ &  2479.888(44)  & 403.243(72) &  16.99(2.89) &	4.7   \\

\hline 
\end{tabular}
\end{table}

\section{Conclusions}
\label{conclusion}

In this paper, we have presented the discovery of two new GW Vir pulsating white dwarfs TIC\,0403800675 and TIC\,1989122424.
We derived atmospheric parameters for TIC\,0403800675 and TIC\,1989122424 by fitting synthetic spectra to the newly obtained low-resolution SOAR/GOODMAN spectra.
The determined spectroscopic parameters demonstrate that TIC\,0403800675 and TIC\,1989122424  are the same to within errors in terms of surface temperature and surface gravity ($T_{\rm eff} = 110,000 \pm 10,000$\,K and $\log g = 7.5\pm0.5$) and they are only different regarding the surface C and He abundance. By performing a fit to the SEDs  and parallaxes we found for both stars radii and luminosities of $R_{\star}=0.019\pm0.002\,R_\odot$ and $\log(L_{\star}/L_\odot)=1.68^{+0.15}_{-0.24}$, respectively. 
Using the \cite{2005A&A...435..631A} and \cite{2006A&A...454..845M} evolutionary tracks of PG~1159 stars, we find a stellar mass for both stars of $0.56\pm0.18 M_{\odot}$ from the $\log g$-$T_\mathrm{eff}$ diagram and  $0.60^{+0.11}_{-0.09} M_{\odot}$ from the HRD.

We have used the TESS 120-second data for both objects, while we made use of 20-second cadence data for only TIC\,0403800675.
Both stars exhibit just two periodicities in their amplitude spectra preventing us to make use of the seismic tools of rotational multiplets and asymptotic period spacing. 
Both pulsational frequencies that we extracted from the light curves are of the order of 7 minutes, attributable to non-radial pulsation $g$-modes.
We also produced sFTs to see if the pulsation modes are resolved and stable throughout the TESS observations, which is limited to a single sector. 
The analyzed data demonstrate that the main pulsational frequencies of our targets are stable and there is no clear pattern for rotational multiplets. 
Unfortunately, the presence of only two oscillation frequencies in the power spectrum of each star prevents us from  making an asteroseismic modelling of these objects. We hope that more periods can be detected in the future, so that we can investigate the internal structure and evolutionary state of these stars through asteroseismological tools. 

GW Vir stars evolve quickly, resulting in a significant period change due to cooling and contraction, which should be detected in a few years.
As a result, these stars can be checked at least once a year to determine evolutionary changes as described by \citet{Costa_Kepler2008}.

\section*{Acknowledgements}

We would like to thank the anonymous referee for the insightful comments and suggestions. 
M.U. thanks CONICYT Doctorado Nacional for financial support in the form of grant number 21190886 and the ESO studentship program.
M.U. thanks Aleksandar Cikota for valuable discussions.
Based on observations performed at the Southern Astrophysical Research (SOAR) telescope as part of the Chilean Time Allocation Committee (CNTAC) program numbers CN2020A-87, CN2020B-74, and CN2021A-52. 
This research uses data from the TESS mission, which was received from the Space Telescope Science Institute's MAST data archive. The NASA Explorer Program provides funding for the TESS program.
This work has made use of data from the European Space Agency (ESA) mission Gaia (\url{https://www.cosmos.esa.int/gaia}), processed by the Gaia Data Processing and Analysis Consortium (DPAC, \url{https://www.cosmos.esa.int/web/gaia/dpac/consortium}). 
Funding for the DPAC has been provided by national institutions, in particular the institutions participating in the Gaia Multilateral Agreement. Part of this work was supported by AGENCIA through the Programa de Modernización Tecnológica BID 1728/OC-AR,and by the PIP 112-200801-00940 grant from CONICET.
This research has made use of NASA's Astrophysics Data System Bibliographic Services, and the SIMBAD database, operated at CDS, Strasbourg, France.

\section*{Data Availability}
Data from TESS is available at the MAST archive \url{https://mast.stsci.edu/search/hst/ui/$#$/}. Ground based data will be shared on reasonable request to the corresponding author.

\bibliographystyle{mnras}
\bibliography{myref.bib} 

\bsp	
\label{lastpage}
\end{document}